\documentclass{sig-alternate-05-2015}
\usepackage{graphicx}
\usepackage{CJKutf8}
\usepackage{color}
\makeatletter
\newif\if@restonecol
\makeatother

\usepackage[ruled,vlined]{algorithm2e}
\begin{document}

\setcounter{secnumdepth}{0}

\setcopyright{acmcopyright}





%

\title{Topical differences between Chinese language Twitter and Sina Weibo}

%
%
%
%
%

\numberofauthors{2} 
\author{
\alignauthor
Qian Zhang\\
       \affaddr{Northeastern University}\\
       \affaddr{Boston, MA, USA}\\
       \email{qi.zhang@neu.edu}
\alignauthor
Bruno Gon\c calves\\
       \affaddr{Center for Data Science}\\
       \affaddr{New York University}\\
       \affaddr{New York, NY, USA}\\
       \email{bgoncalves@gmail.com}
}

\maketitle
\begin{abstract}
Sina Weibo, China's most popular microblogging platform, is currently used by over $500M$ users and is considered to be a proxy of Chinese social life. In this study, we contrast the discussions occurring on Sina Weibo and on Chinese language Twitter in order to observe two different strands of Chinese culture: people within China who use Sina Weibo with its government imposed restrictions and those outside that are free to speak completely anonymously. We first propose a simple ad-hoc algorithm to identify topics of Tweets and Weibo. Different from previous works on micro-message topic detection, our algorithm considers topics of the same contents but with different \#tags. Our algorithm can also detect topics for Tweets and Weibos without any \#tags. Using a large corpus of Weibo and Chinese language tweets, covering the period from January $1$ to December $31$, $2012$, we obtain a list of topics using clustered \#tags that we can then use to compare the two platforms.  Surprisingly, we find that there are no common entries among the Top $100$ most popular topics. Furthermore, only $9.2\%$ of tweets correspond to the Top $1000$ topics on Sina Weibo platform, and conversely only $4.4\%$ of weibos were found to discuss the most popular Twitter topics. Our results reveal significant differences in social attention on the two platforms, with most popular topics on Sina Weibo relating to entertainment while most tweets corresponded to cultural or political contents that is practically non existent in Sina Weibo. 
\end{abstract}

%
%


%
%

%
%



\section{Introduction}\label{intro}

China is known for its rich internal Internet ecosystem where Chinese alternatives to most foreign Internet services flourish. This is due not only to cultural differences that prevent foreign websites from gaining a large market share, but also due to stringent government controls that sometimes prevent foreign Internet companies from selling their services or that outright block access to them. 

Sina Weibo, as China’s most popular microblogging platform, is perhaps the most visible face of China's own internal version of the Internet. It is currently used by over $500M$ users and, similarly to its foreign counterpart Twitter~\cite{gao2012comparative} that is widely considered to be a proxy for its users social life and interests~\cite{bamman14-1,phelan09-1,ciulla12-1,metaxas11-1}, it has recently started to draw the attention of researchers everywhere~\cite{Rauchfleisch2014-1,fu2013assessing,fu2013reality,bamman2012censorship}. 

Sina Weibos origins date back to $2009$ but it wasn't until $2011$ that it rose to prominence. Since July 2009, Twitter has been blocked in China \cite{bamman2012censorship}, leaving national alternatives such as Sina Weibo as the only alternative. In March 2012, Weibo started requiring its users to associate their profile with their true identity~\cite{Rauchfleisch2014-1} while still giving users the option to display whichever screenname they wished. 

The previous works on topic detection on microblogs are usually designed for pre-selected specific topics \cite{phuvipadawat2010breaking,li2013improved} or only for short-messages with \#tags \cite{Tsur2013}. However, the majority of Tweets are not \# tagged \cite{phuvipadawat2010breaking}, and there is few work focusing on automatic topic detection for microblogs. We first propose a simple ad-hoc algorithm to identify topics on microblogs without pre-selection, and cluster those microblogs without \#tag into the detected topics. More importantly, without the assumption that a \#tag represents a unique topic, our algorithm merges microblogs of the same contents but with different \#tags. 

Past studies on Chinese microblogging platforms \cite{fu2013assessing,bamman2012censorship} mainly focused on censorship and analyzed deleting practices on microblogs containing censored key words. Others compared the user behaviors, texture features of posts and temporal dynamics of re-posting \cite{gao2012comparative} and an artificially selected categorical events \cite{Shuai2014} on Sina Weibo and Twitter. There is little research on comparing the collective attention of Chinese microbloggers in a large scale. Here, we take a first step in this direction by proposing an algorithm to model and compare Sina Weibo and Twitter. By contrasting the discussions occurring on these two platforms, we can observe two different versions of chinese culture: people inside China ($94.8\%$ of geotagged Weibos are within China) and those outside ($93.7\%$ of geotagged Tweets are located outside China). Despite China's growing global relevance, and due to the complexity of its language, the number of people outside China learning Chinese as a second language is still very small. A recent study estimates that, in $2009$, just over $60,000$ students in American universities were taking Chinese language classes, compared with over $865,000$ studying Spanish~\cite{report}, indicating that people who Tweet in Chinese outside of China are likely either Chinese expats or from Chinese heritage. While some analyses have been performed on geographically distributed populations speaking the same language~ \cite{goncalves14-1}, this combination of technically equivalent services serving populations with a similar cultural background that are isolated from each other is unique and provides us with the perfect opportunity to study the cultural differences in the virtual world between Chinese speakers inside and outside China.

In summary, we perform a topical comparison of both Twitter and Sina Weibo. Our results reveal significant differences in social attention distribution across both platforms, with the most popular topics on Sina Weibo relating to entertainment while the most topics in Twitter corresponded to cultural or political contents.

\section{Data Description}
\label{dataset}
We use the dataset of Sina Weibo from Open Weiboscope Data Access \cite{fu2013assessing,fu2013reality}. The dataset contains $226.8$ million Weibo posts (Weibo for short) collected over the full course of $2012$. The Twitter dataset used in this study was extracted from the raw Gardenhose feed \cite{ratkiewicz2011truthy}, an unbiased sample of $10\%$ of the entire Twitter dataset that provides a statistically significant real time view of all Twitter account activity \cite{TwitterAPI}. To identify Tweets and Weibo in Chinese language, we perform language detection using the ``Chromium Compact Language Detector'' \cite{ChromeLanguage}. See \cite{mocanu2013twitter} for further details. This way, we collected $12.3$ Million Tweets and $216.8$ Million Weibo in both simplified and traditional Chinese language covering the entire year of $2012$. The Sina Weibo dataset also include microblogs which are not accessible to the public, either censored or self deleted. Following \cite{fu2013assessing}, we consider weibos deleted by the censorship (with message ``permission denied'' from API). In total, we considered $74,132$ deleted weibos for our study.

\section{Clustering microblogs into topics}\label{algorithm}

Topic modeling on micro-messages is still challenging due to its inherent sparseness \cite{Tsur2013} and noise \cite{li2013improved}. In this study, we take microblogs with \#tags as potential known topics. There are $2.06M$ tweets and $18.9M$ weibos with \#tags. We build a vocabulary vector space on each microblogging platform with words of high-frequency (high TF-IDF score), and cluster similar \#tags into a specific topic. For instance, for the Top $100$ \#tags on Sina Weibo and Twitter, we merge them into $83$ and $58$ topics respectively. For the rest of microblogs without \#tags, we assign them to topics that are closest to them in the vocabulary vector space. To reduce statistical fluctuations we restrict our study to the Top $1000$ topics in each platform. In total, $\sim 20.8\%$ weibos are classified into popular topics on Weibo and $\sim 34.5\%$ of tweets discuss popular topics on Twitter. In the remaining part of this section, we briefly describe our algorithm of clustering microblogs into topics.

\textbf{Preprocessing.}
We first filter microblogs by removing the words representing short URLs and mentioning other users (``@username''). Filtered microblogs in traditional Chinese are then converted to simplified Chinese with the python-jianfan library \cite{jianfan}. Chinese word segmentation is performed using Jieba \cite{jieba} and part-of-speech tagging (POS) is performed  following \cite{li2013improved}. This way each microblog is represented as a set of words tagged as noun, name, location, organization, time, place word, position word or verb.

\textbf{Vector representation.}
We merge all microblogs with the same \#tag $h_i$ as a document $D_{h_i}$, and calculate its TF-IDF (term frequency–inverse document frequency). For each $D_{h_i}$, we exclude the words with length less than 2 since a single character word in Chinese can be noisy and under-representative, and choose the first $10$ words $t_{h_i}^{j}, j=0\ldots 9$ with highest frequency and their TF-IDF weights $w_{h_i}^{j}$, and its vocabulary vector can be written as $V_{h_i} = (0:0,...,0:0, t_{h_i}^{0}:w_{h_i}^{0}, t_{h_i}^{1}:w_{h_i}^{1}, ..., t_{h_i}^{9}:w_{h_i}^{9},0:0,...,0:0)$, and $|V_{h_i}|=10N$ where $N$ is the number of \#tags we select.

\begin{figure}[ht]
\centering
\includegraphics[width=.99\columnwidth]{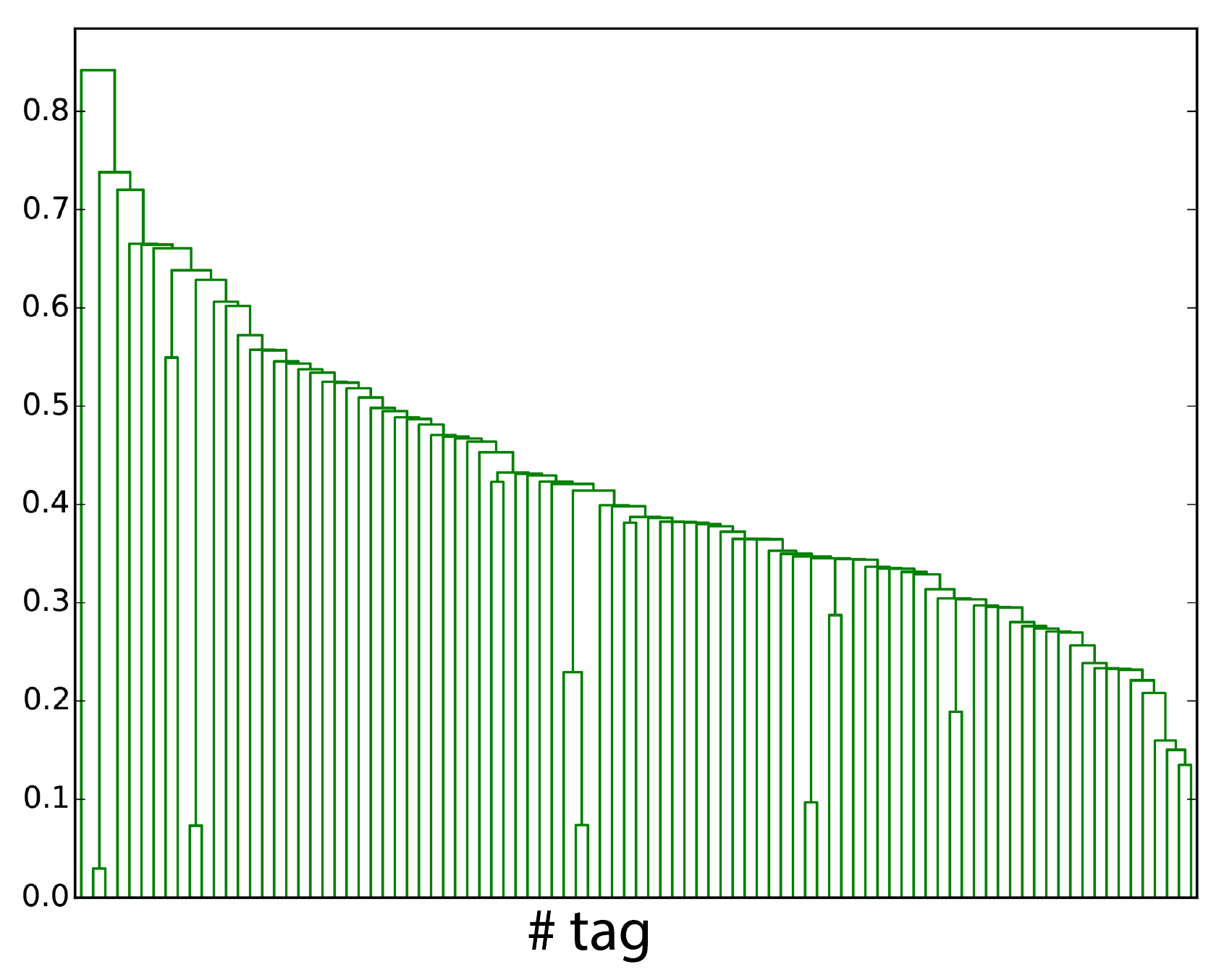}
\caption{\textbf{Cluster Dendrogram}. Cluster dendrogram for Top $100$ \#tags on Sina Weibo contains a number of simplicifolious clades, indicating most of topics labelled by \#tags are distant from each other.}\label{fig:clustering}
\end{figure}

\textbf{Clustering.}
Since several similar \#tags likely refer to the same topic, we further cluster \#tags into topics using hierarchical clustering. In Figure \ref{fig:clustering}, we show the dendrogram for the Top $100$ \#tags on Sina Weibo platform, calculated using cosine distances in the embedding vector space  $d_{h_i, h_j}=V_{h_i}\cdot V_{h_j}/\|V_{h_i}\|\|V_{h_j}\|$. Interestingly, most clades are simplicifolious, indicating that distribution of words for each \#tag is substantially different from the distribution in others. We observe similar dendrogram for Top $100$ \#tags on Twitter (figure not shown). Thus, we apply a modified divisive clustering method (Algorithm \ref{algo1}), where we iteratively divide the largest cluster into a small cluster and a large one, until the size of the small cluster is $0$.
\begin{algorithm}\label{algo1}
 \KwIn{The hierarchical cluster $L$}
 \KwOut{The set of topics $S_T$ }
 initialization\;
 -- $S_T = \emptyset$ \;
 \While{$L \neq \emptyset$}{
 Partition $L$ into a larger cluster $B_L$ and a small cluster $B_S$\;
 $L \leftarrow B_L$\;
 $S_T \leftarrow S_T \bigcup \{B_S\} $\;
}
 \caption{Merging \#tags into topics}
\end{algorithm}

After merging \# tags into topics, each topic $t_i$ in vocabulary vector space now is defined as  $V_{t_i} = (0:0,...,0:0, t_{t_i}^{0}:w_{t_i}^{0}, t_{t_i}^{1}:w_{t_i}^{1}, ..., t_{t_i}^{9}:w_{t_i}^{9},0:0,...,0:0)$, and $|V_{t_i}|=10T, T\leq N$.  $t_{t_i}^{j}, j=0\ldots 9$ is now the first $10$ words with highest frequency in a topic $t_i$ and their TF-IDF weights $w_{t_i}^{j}$. The centroids of the final clusters are taken to represent topics in the vector space of each platform. To classify the remaining microblogs on one platform, we measure the cosine distance between the centroid of a topic $t_j$, and each microblog $m$, $d_{t_j, m}$. If $d_{t_j, m}$ is smaller than a threshold $d_t$, we consider the microblog $m$ is discussing the topic $t_j$, shown in Algorithm \ref{algo2}.
\begin{algorithm}\label{algo2}
 \KwIn{The set of topics $S_T$ and a set of microblogs $M$}
 \ForEach{topic $t_j$ in $S_T$}{
computing the centroid $C_{t_j}$ in the vocabulary vector space where $|C_{t_j}| = 10T$ \;
label $t_j$ with its \# tags clustered from Algorithm \ref{algo1}\;
 }
 \ForEach{microblog $m$ in $M$}{
 \If{$m$ contains \# tag $h_x$}{
 Assign $m$ into the topic $t$ with \# tag $h_x$\;
 }
\Else{
set $V_{m} = (t_{t_0}^{j}:w_{t_0}^{j},... t_{t_i}^{j}:w_{t_i}^{j}, ..., t_{t_{T-1}}^{j}:w_{t_{T-1}}^{j})$, where $j=0\ldots 9$ \;
let $T_m = None$ be the topic for $m$\;
let $D_m^{min} = \infty$ be the minimum distance from $m$ to any centroid of clustered topic \;
 \ForEach{topic $t_j$ in $S_T$}{
	computing the distance $d_{t_j, m} = V_{m}\cdot C_{t_j}/\|V_{m}\|\|C_{t_j}\|$ \;
	\If{$d_{t_j, m} < d_t$}{
	$D_m^{min} \leftarrow d_{t_j, m}$\;
	$T_m \leftarrow t_j$\;
	}
 }
 \If{$T_m$ is not None}{
 Assign the microblog $m$ into the topic $T_m$ \;
 }
 \Else{
 Tag the microblog $m$ to be `unknown topic' \;
 }
}
}
 \caption{Clustering micrblogs into topics}
\end{algorithm}

To determine the threshold $d_t$, we measure the distribution of distance between a centroid of a topic and microblogs inside that topic. About about $65\%$ of tweets and $76\%$ of weibos have distances less than $0.9$ to their topical centroid. Meanwhile, if we measure distances from a microblog to centroids of other topics, on average only about $9.2\%$ of microblogs outside a topical centroid have distances less than $0.9$. Therefore, we use $d_t=0.9$ as our threshold.

\section{Results and Discussion}
\begin{figure}[ht]
\centering
\includegraphics[width=.99\columnwidth]{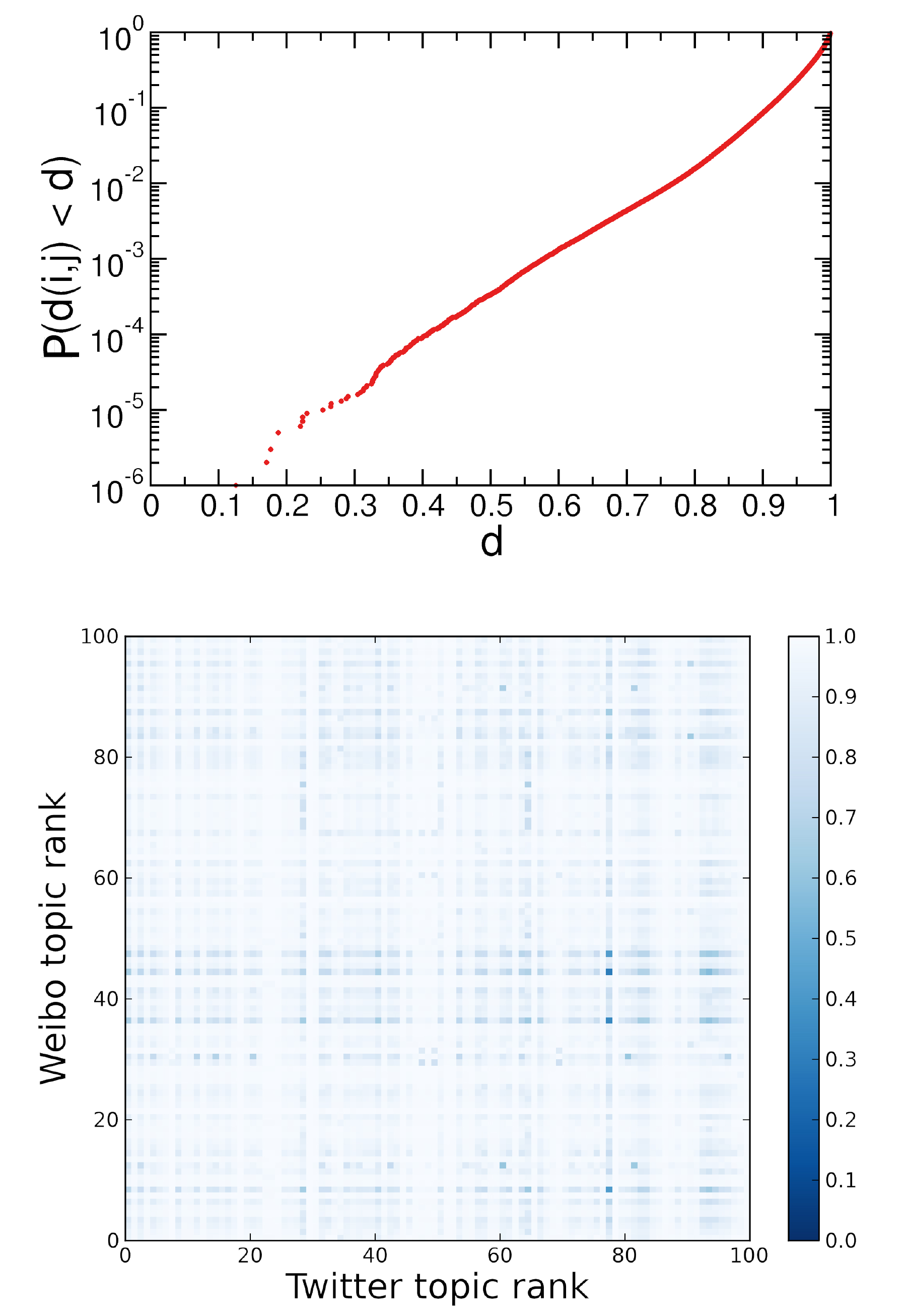}
\caption{\textbf{Distance between popular topics on Twitter and Sina Weibo}. (A) The cumulative distribution function of the distance $d_{ij}$ as the cosine distance between the centroid of topic $i$ in Top $1,000$ topics on Twitter and topic $j$ in Top $1,000$ topics on Sina Weibo. (B) The distance matrix between the centroid of topic $i$ in Top $100$ topics on Twitter and topic $j$ in Top $100$ topics on Sina Weibo. The distance $0$ refers two exactly similar topics, and $1$ indicates that two topics are completely different.}\label{fig:distacematrix}
\end{figure}

Our analysis aims to compare topical spaces in Chinese language on different microblogging systems. With identified centroids in the vocabulary vector space defined in the last section, we first calculate the distance between the centroids of the Top $1000$ topics on the two platforms. We define $d_{ij}$ as the cosine distance in the vocabulary vector space between the centroid of topic $i$ on Twitter and topic $j$ on Sina Weibo. Figure \ref{fig:distacematrix}-A shows the cumulative distribution function of distance $d_{ij}$ for $10^3\times 10^3$ pairs of topical centroids. Surprisingly, only $8.9\%$ pairs of topics have distance less than $0.9$. In Figure \ref{fig:distacematrix}-B, we show the distance between Top $100$ topics on Weibo and Top $100$ topics on Twitter. Surprisingly, the distance between $91\%$ of pairs of Top $100$ topics on the two platforms is larger than $0.9$, indicating that microbloggers in each platform have significantly different conversation topics and interests.

In Table~\ref{tb:top10topics}, we provide the Top $10$ topics in Chinese language on Sina Weibo and Twitter to illustrate the differences. On Sina Weibo, $\sim 1.88\%$ of entire datasets can be classified into top 10 topics ($\sim 9.01\%$ for the Top $100$); while on Twitter $\sim 13.43\%$ over all tweets are categorized into top 10 topics ($\sim 20.8\%$ for Top $100$). The microblogs on Sina Weibo focus on entertainment (singers, actors and games) and advertising. In contrast, on Twitter, there is no commercial advertisement appearing, and the last two topics are about games. The others are all corresponding to political contents.

\begin{CJK*}{UTF8}{gbsn}
\begin{table*}[ht]
\centering
\begin{tabular}{l| c c r | c c r}
\hline
&&Topic&&&Topic&\\
 Rank & Sina Weibo & in English & \%  & Twitter & in English & \%  \\ [0.5ex] 
\hline
1 & 三国来了 & an online game & 0.51 & 陈光诚 & \begin{tabular}{@{}c@{}}Chen Guangcheng \\(a Chinese civil rights activist)\end{tabular} & 3.56 \\[12pt]
2 & 林峰 & \begin{tabular}{@{}c@{}}Raymond Lam \\(a singer from Hongkong)\end{tabular}  & 0.39 & 乌坎 & Wukan protest & 2.56 \\[15pt]
3 & 晚安/早安 & good morning/night & 0.38 & Freetibet & Free Tibet & 1.62 \\[12pt]

4 & 微博客户端 & Sina Weibo app & 0.36 & 李旺阳 & \begin{tabular}{@{}c@{}}Li Wangyang (a Chinese \\dissident labor rights activist)\end{tabular} & 0.09 \\[12pt]

5 & 搞笑 & joke & 0.07 & 温云 & @wenyunch & 0.97 \\[12pt]

6 & 美图秀秀 & \begin{tabular}{@{}c@{}} Meitu (an iOS/Android \\app to edit pictures)\end{tabular} & 0.04 & 抗暴 & Tibetan Uprising Day & 0.88 \\[12pt]

7 & 有奖转发 & re-posting to win a prize & 0.04 & 达赖喇嘛 & Dalai Lama & 0.68 \\[12pt]

8 & WeicoLomo & \begin{tabular}{@{}c@{}}An iOS/Android app for\\ Weibo to record video\end{tabular} & 0.04 & 钓鱼岛 & \begin{tabular}{@{}c@{}}Uotsuri Jima \\Diaoyu Dao / Diaoyutai \end{tabular}& 0.66 \\[12pt]

9 & 韩庚 & \begin{tabular}{@{}c@{}}Han Geng \\(a Chinese singer and actor)\end{tabular} & 0.03 & ipadgame & iPad game & 0.61 \\[12pt]

10 & 新版微博 & new version of Sina Weibo & 0.02 & 武士朝代 & an Andorid game & 0.48 \\
\hline
\end{tabular}
\caption{\textbf{Top 10 Topics on Sina Weibo and Chinese language Twitter in $2012$.}}\label{tb:top10topics} 
\end{table*}
\end{CJK*}

In the previous section, we have classified weibos and tweets into the topical space in their own vocabulary vector space. For an unclassified weibo or tweet, we calculate its distance to centroids on both platforms, and assign it to the closest topic. Interestingly, we find there are only $\sim 9.2\%$ of tweets correspond to the Top $1000$ topics on Sina Weibo platform, and only $\sim 4.4\%$ of weibos were discussing the most popular topics on Twitter. Chinese microbloggers speaking the same languages on two platforms share a few social attentions.

We further investigate deleted weibos that were likely censored \cite{fu2013assessing} by checking if they belong to topics which appear on Twitter. In total, $1,558$ deleted weibos can be classified into the Top $100$ topics on Twitter. We re-rank the topics in accordance with the frequency of deleted weibos. The Kendall rank correlation coefficient between the top $100$ topics for all tweets and for deleted weibos is $\tau = 0.31$, with $p$-value $4\times 10^{-6}$. In Table \ref{tb:top10topicsdeleted}, we list Top $10$ topics for deleted weibos on Twitter's vector space. Compared with popular topics on Sina Weibo, the deleted weibos are significanlty more likely to discuss political issues.
\begin{CJK*}{UTF8}{gbsn}
\begin{table}[ht]
\centering
\begin{tabular}{l| c c }
\hline
 rank & topic on Twitter & in English   \\ [0.5ex] 
 \hline
 1 & fb & Facebook \\
 2 & 乌坎 & Wukan protest \\
 3 & np & Now Playing \\
 4 & hitbag & - \\
 5 & AutoShare & - \\
 6 & GFW & Great Firewall\\
 7 & HK71 & Hong Kong 1 July march \\
 8 & bot & - \\
 9 & A片 & Adult movie \\
 10 & JapanLife & - \\
\hline
\end{tabular}
\caption{\textbf{Top 10 topics in deleted weibos on Twitter's vector space .}}\label{tb:top10topicsdeleted} 
\end{table}
\end{CJK*}

\section{Conclusion}
The social attention of online users from the same cultural backgrounds but living in different countries might be different due to the changes of social environments. In this study, we take the first steps toward understanding such differences.

Sina Weibo is used almost exclusively within China while most Chinese language use of Twitter occurs almost exclusively outside Chinese borders. By comparing the most popular topics in these two platforms we can, for the first time, observe how the interests of two populations, with similar cultural backgrounds, differ. Surprisingly, we find that there is very little overlap between the two attention profiles. Weibo users speak mostly about popular culture and games while Twitter users focus mostly on political issues. 

The reasons behind this divergence are hard to discern but can likely be attributed to one of two factors: lack of interest for political topics within China or a high degree of self-censorship that prevents Chinese from discussing politics in public. A small indication towards this second hypothesis is the list of topics seen in deleted Weibos (see Table~\ref{tb:top10topicsdeleted}) that have higher political content. It is worth to remark that our algorithm of detecting topics still depends on \# tags, and some of such \# tags may not necessarily be a social topic but likely represent some commercial web/mobile applications. Manual annotations could be included in the future work to improve the topic detection results. Another key datapoint we are missing to fully clarify this question is the number of people who use foreign VPN services as a way of being able to reach Twitter where the discussion is more politically centered. An analysis of this interesting factor will be the subject of future study. The proposed methodology in this paper can be easily applied to any other languages across different online conversation platforms if data are available.

Another possibility worth considering when comparing user behavior across two different platforms are the technical differences between the platforms are not to be excluded. However, the similarity between the two platforms likely minimizes this effect. Indeed, it would  be difficult to argue that Twitter is, on technical grounds, any more or less suitable to discussion of the topics listed on the right side of Table $1$ than Sina Weibo or vice-versa. A final possibility is simply that different "cultural norms" have emerged in the two platforms~\cite{zhou2011understanding} with the Sina Weibo community naturally becoming much more focused on pop-culture and entertainment and Twitter becoming more political. 


	

\end{document}